\documentclass[twocolumn,times,final]{elsarticle}

\usepackage{prletters}

\usepackage{amssymb,amsmath,mathtools}
\usepackage{ulem}
\usepackage{wrapfig}
\usepackage{bbm}
\usepackage{graphicx,tikz}
\usepackage{psfrag}
\usepackage{xcolor}
\usepackage{siunitx}
\usepackage{tabularx}
\usepackage{multirow,array,booktabs}
\usepackage{caption}
\usepackage{subcaption}
\usepackage{wrapfig}
\usepackage{upgreek}
\usepackage{float}

\biboptions{sort&compress}

\usepackage{algorithm}
\usepackage{algpseudocode}

\newtheorem{problem}{Problem}

\newcommand{\expect}[1]{\mathbb{E}_{#1}}

\journal{Pattern Recognition Letters}

\usepackage{lineno}
\usepackage{hyperref}
\modulolinenumbers[5]

\begin{document}

\begin{frontmatter}

\title{Information-Theoretic Policy Learning from Partial Observations \\ with Fully Informed Decision Makers}

\author[mymainaddress,mysecondaryaddress]{Tom Lefebvre}
\address[mymainaddress]{T. L. is with the Dynamic Design Lab (D\textsuperscript{2}LAB) from the Departement of Electromechanical, Systems and Metal Engineering, Faculty of Engineering and Architecture, Ghent University, Technologiepark 131, 9052 Zwijnaarde, Belgium}
\address[mysecondaryaddress]{Corresponding author: \href{tom.lefebvre@ugent.be}{tom.lefebvre@ugent.be}.}
 
\begin{abstract} 
{In this work we formulate and treat an extension of the Imitation from Observations problem}. Imitation from Observations is a generalisation of the well-known Imitation Learning problem where state-only demonstrations are considered. In our treatment we extend the scope of Imitation from Observations to feature-only demonstrations which could arguably be described as partial observations. Therewith we mean that the full state of the decision makers is unknown and imitation must take place on the basis of a limited set of features. We set out for methods that extract an executable policy directly from those features which, in the literature, would be referred to as Behavioural Cloning methods. Our treatment combines elements from probability and information theory and draws connections with entropy regularized Markov Decision Processes.
\end{abstract}

\end{frontmatter}


\section{Introduction}

Imitation Learning (IL) refers to the process by which a student tries to learn how to execute a task by observing an experienced teacher demonstrate the task. Usually the student is allowed to collect a number of those demonstrations before attempting the task \cite{argall2009survey,osa2018algorithmic,schaal1996learning}. IL is known to be useful to treat hard control problems such as driving \cite{pan2020imitation} and grasping \cite{zhang2018deep,jiang2021manipulator,zhang2019effective}. 

Conventionally it is assumed that the student has gained access to demonstrations that include both the teacher's \textit{states} $\{x_t\}_t$ (e.g. position, velocities) as well as its \textit{actions} $\{u_t\}_t$ (e.g. forces). By definition this restrictive problem statement rules out a number of potentially useful teachers solely because said {actions} cannot be accessed. Though, even more restrictive is the assumption that the demonstrations must relate to the student performing the task rather than the teacher since this suggests that the states and actions must be those of the student and not those of the teacher (e.g. a human demonstrating a task by grabbing the end-effector whilst the robot collects measurements from its joints). The latter is known as the \textit{embodiment mismatch} problem \cite{sermanet2018time,torabi2019recent}. Alternatively one might imagine scenario's where the students has mere access to a sequence of features, $\{z_t\}_t$, representative of the teacher's demonstration, and cannot access it's own states and actions.
\newpage

To accommodate these flaws, a smaller number of studies has began to pose and treat IL problems with \textit{state-only} demonstrations. Here clearly only the teacher's \textit{states} can be accessed \cite{sharma2019third,torabi2018bc,liu2018imitation,sun2019provably}. It is argued that the so called \textit{Imitation from Observation} (IfO) paradigm offers a more natural way to consider learning from a teacher, and exhibits more similarity with the way many biological agents appear to approach imitation \cite{torabi2019recent}. However, as is implied by the term \textit{state-only} demonstrations, many recent studies that treat the IfO problem still assume full state observability, i.e. $z_t = x_t$. {This setting is arguably evenly restrictive as full IL, given how rare a teacher and a student are that have the exact same state. Think of a human demonstrating a trajectory to a robot without grasping the robot.} 

In this work we consider the setting where only a limited set of representative features {or partial observations} are available. {We refer to this problem as the Imitation learning from Partial Observations (IfPO) problem.} {In the context of IfPO, the goal is for the teacher and student to have the same effect on the environment, rather than the student doing the exact same thing as its teacher. In the setting of IfPO, the goal is for the student to showcase behaviour that appears similar to the behaviour displayed by the teacher to an \textit{objective observer} that is merely interested in those features that were collected in the first place. }

{Our methodological contributions are the following
	\begin{enumerate}
		\item First we formalize the IfPO problem {and put forth a probabilistic model for how the students dynamics may spawn the set of features displayed by the teacher.}
		\item {We formulate a straightforward treatment of the IfPO problem by recasting it as a Bayesian inference problem which in turn leads us to derive a first learning algorithm. }
		\item Second, {we recast the solution of the Bayesian inference as an information-theoretic projection. This reformulation allows us to generalise the treatment to multiple measurement sequences and to propose a sibling strategy by means of the reciprocal information-theoretic projection.}
\end{enumerate}

These aspects describe a first step towards a generic treatment of the IfPO problem and an extension to more general models.} {Then because the resulting policy learning algorithms demonstrate great similarity with the solution of Markov Decision Processes (MPDs), we pursue that intuition to some extent. Finally we specialize the concepts to linear-Gaussian systems.} {A main restriction of the proposed methods is the availability of the student's emission model (see later). Future work may focus on learning the emission model too but will likely require to embed the proposed approached in a more general framework.}


\begin{figure}
	\centering
	\includegraphics[scale=1]{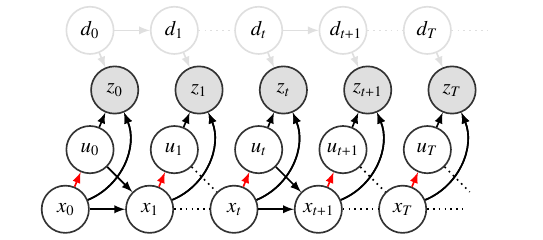}
	\caption{Graphical representation of the imitation learning problem with partial observations and a fully informed decision maker. }
	\label{fig:CHMM}
	\vspace*{-24pt}
\end{figure}

\section{Problem formulation}

As a proxy for the student's imitation learning process, we propose the probabilistic graph model in Fig. \ref{fig:CHMM}
\begin{equation*}
\begin{aligned}
x_t &\sim p(x_{t+1}|x_t,u_t) = \tau_t \\
u_t &\sim p(u_t|x_t) = \rho_t \\
z_t &\sim p(z_t|x_t,u_t) = \epsilon_t
\end{aligned}
\end{equation*}
Let us here briefly verbalize our notation. Any of the spaces $\mathcal{X}$, $\mathcal{U}$ or $\mathcal{Z}$ can be discrete or continuous. For brevity we introduce the variable tuple ${\xi} = ({x},{u})$ and the formatting $\underline{X}_t = \{x_0,\dots,x_t\}$ and $\overline{X}_t = \{x_t,\dots,x_T\}$ for leading or trailing subsequences. Throughout we use $\mathcal{P}$ to refer to \textit{sets} of probability density functions. The set's arguments are implied by context.
\begin{enumerate}
	\item $x_t\in\mathcal{X}$ represents the student's Markovian state, $\mathcal{X}$ denotes the state space. The probabilistic transition dynamics are governed by the (time-varying) functions $\{\tau_t\}_t$.
	\item $u_t\in\mathcal{U}$ denotes the student's action, $\mathcal{U}$ denotes the action space. We assume that prior to the learning process the student has access to some auxiliary policy $\rho_t$.
	\item $z_t\in\mathcal{Z}$ represent features of the student's shared with the teacher. $\mathcal{Z}$ denotes the feature space, the (time-varying) observation or emission model is given by $\{\epsilon_t\}_t$.
	\item $d_t$ represents the latent state of the teacher. {The representation of the teacher's state is strictly demonstrative and included as a justification for the origin of the feature sequence. However it is irrelevant what the teacher's exact state is and therefore also irrelevant what the teacher's transition and emission model are. Think of a human demonstrating a reference trajectory with its arm. We are unaware of the teacher's state nor its reasons to showcase that trajectory, yet we can measure the hand's trajectory.} 
\end{enumerate}

We deliberately do not model the student's actions, $\{u_t\}_t$, to be the result of an MPD (see sec. \ref{sec:connections-with-optimal-control}). In general there is no reason to assume that the teacher acted rationally, i.e. acted so to optimize some utility function, {so there is no reason for the student attempting to minimize some utility function}. Furthermore this would imply that the {student} acts deterministically which we can also not know for sure. The main conceptual assumption is that the student and the teacher share the feature sequence $\underline{Z}_T$. So even though the feature sequence spawned from the dynamics of the teacher, we assume that with equal probability they could have spawned from the student. {In this preliminary investigation, we also assume that the \textit{proprioceptive} subsystem of the student is perfect and we have access to a perfect state estimate. Thus, the student's policy is governed by a probability density function conditioned on it's state $x_t$. We refer to such a decision maker as being \textit{fully informed}.}


As such we consider the following problem
\begin{problem}
\label{prob:2}
Learn policy functions $\{\pi_t\}_t$ provided that the decision maker is fully informed. Assume that the transition $\{\tau_t\}_t$ and emission $\{\epsilon_t\}_t$ models are known as well as multiple feature sequences $\{z_t^n\}_{t,n}$, representative for the dynamic behaviour that the decision maker's policy ought to induce. 
\end{problem}

{
	By definition the strategies derived in this work are model-based. Our reasoning is that although the student might not know how to perform a task, it is reasonable to assume it has knowledge about its own transition model. {For now we also assume that student has access to its emission model, i.e. a probabilistic mapping from state and actions to the relevant features. }For example, the robot is aware of its own forward kinematics. 
}

\section{Algorithms for policy learning} 
We propose two, as far as we are aware of, original strategies to solve problem \ref{prob:2}. In what follows we will refer to any of the proposed policy extraction strategies as an A Posteriori Control Distribution (APCD). Provided that we do not model the decision process as an MDP and that we assume to have access to a model, the strategies would classify as model-based BC algorithms. Though as mentioned earlier, the computational procedures are closely related to those characteristic for MDPs. The exact relation between the proposed BC algorithms and IRL methods will be investigated to some extent in sec. \ref{sec:connections-with-optimal-control}. {An important feature of the proposed methods is the ability to assimilate the information contained by multiple demonstrations.}

The first strategy to derive an APCD follows a standard Bayesian argument. As a result, the concept only applies to a single measurement sequence. We investigate the solution's computational properties and whilst doing so explore whether the same result may be expressed as an optimization problem {that generalizes over multiple feature sequences}. The corresponding optimization problem turns out to be related to the information-theoretic Moment or M-projection. 
We refer to this APCD as the {\textit{vanilla}} APCD (V-APCD). The terminology anticipates another APCD which we shall refer to as the \textit{\textit{natural}} APCD (N-APCD). The class of N-APCDs is obtained by considering and solving the reciprocal information-theoretic Information- or I-projection and is developed further in sec. \ref{sec:natural-apcd}. Our use of terminology is inspired by \textit{vanilla} and \textit{natural} gradients in policy search where the natural gradient follows a relative entropy constraint corresponding the I-projection \cite{pierrot2021}.


\subsection{A primer on Information-Theoretic projections}
Probabilistic inference refers to the process of reasoning with incomplete information according to rational principles. Inference principles therefore determine how to update a prior belief into a posterior belief when new information becomes available. Bayesian inference can be used to process information that is represented by the outcome of experiments, i.e. empirical evidence. The Information- and Moment-projection, respectively abbreviated to the I- and M-projection, are information-theoretic concepts that can be used to process information represented by constraints that affect the belief space. We could arguably refer to such evidence as \textit{structural}. 

The concepts are based on the relative entropy $\mathbb{D}[\pi\parallel\rho] = \expect{\pi}[\log \frac{\pi}{\rho}]$ between distributions $\pi\in\mathcal{P}$ and $\rho\in\mathcal{P}$. The relative entropy is a measure of the inefficiency of assuming	that the distribution is $\rho$ when the true distribution is $\pi$ \cite{cover2006elements}. According to the principle of maximum entropy advocated by Jaynes amongst others, the relative entropy should be minimized if we want to encode some form of new information (usually an expectation, i.e. $\expect{\pi}[f] = \mu$) into the prior $\rho$. 
\begin{itemize}
	\item\textbf{I-projection} The I-projection and maximum entropy principle are equivalent with $\pi^\bullet$ the I-projection of $\rho$ onto $\mathcal{P}^\bullet$.
	\begin{equation*}
	\pi^\bullet = \arg\min\nolimits_{\pi\in\mathcal{P}^\bullet} \mathbb{D}[\pi\parallel\rho] 
	\end{equation*}
	\item\textbf{M-projection} The M-projection is the reciprocal of the I-projection with $\pi^\star$ the M-projection of $\rho$ onto $\mathcal{P}^*$.
	\begin{equation*}
	\pi^\star=\arg\min\nolimits_{\pi\in\mathcal{P}^*} \mathbb{D}[\rho\parallel\pi] 
	\end{equation*}
\end{itemize}

The set $\mathcal{P}^*\subset\mathcal{P}$ represents the constrained belief space. The relative entropy is a divergence and not a distance and thus asymmetric in its arguments. Therefore the I-projection and the M-projection do not yield the same projection \cite{bishop2006pattern,pml1Book}. They are either \textit{zero forcing} or \textit{avoiding} for $\pi$. As a result the I-projection will underestimate the support of $\rho$ and vice versa.

\subsection{Vanilla APCD}\label{sec:vanilla-apcd}

First we describe the idea of the V-APCD for a single measurement sequence, $\underline{Z}_T$. Then we generalise the result to multiple measurement sequences, $\{\underline{Z}_T^n\}_n$, by reformulating the solution as an information-theoretic projection.

\subsubsection{Bayesian argument}\label{sec:bayesian-approach} 

Reconsidering the interpretation of problem \ref{prob:2} we propose to synthesise the V-APCDs, $\{\pi_t^\star\}_t$, simply by conditioning the posterior marginal, $p(\xi_t|\underline{Z}_T)$, on the state, i.e. $p(u_t|x_t,\underline{Z}_T)$. This approach is closely related to BC algorithms, given that first we infer the most likely state-action distribution for given measurements and than infer the most likely policy. The ML community has developed various methods to practice inference on probabilistic graph models and hence the probabilities $p(\xi_t,\underline{Z}_T)$ and $p(u_t|x_t,\underline{Z}_T)$ could be calculated using one of these methods \cite{sarkka2013bayesian}. However such a general purpose method would not fully exploit the computational structure of the model nor would it yield further insights in the problem at hand. 
\begin{equation*}
\pi^\star_t({x}_t) = p({u}_t|{x}_t,\overline{Z}_{T})
\end{equation*}

As a result of the Markov assumptions we have that the posterior marginal must be equivalent to $p({u}_t|{x}_t,\overline{Z}_t)$. That is we can drop the leading subsequence $\underline{Z}_{t-1}$ from $\underline{Z}_T$. This observation resonates with the common sense that once we arrive at some state, ${x}_t$, we can only hope to reproduce measurements $\overline{Z}_t$ but can no longer hope to affect the preceding measurements $\underline{Z}_{t-1}$. Thus we use Bayes' rule to decompose the V-APCD as follows. 
\begin{equation*}
p(u_t|x_t,\overline{Z}_t) = \rho_t(u_t|x_t) \frac{p(\overline{Z}_t|\xi_t)}{p(\overline{Z}_t|x_t)}
\end{equation*}
The decomposition reduces the problem to finding efficient expressions for the probabilities $p(\overline{Z}_t|x_t)$ and $p(\overline{Z}_t|\xi_t)$. The latter can be recognized as a generalisation of the backward filtering distribution. The former can be derived from there.	
\begin{equation*}
p(\overline{Z}_t|x_t) = \int \rho_t(u_t|x_t) p(\overline{Z}_t|\xi_t)\text{d}u_t = \expect{\rho_t(u_t|x_t)}\left[p(\overline{Z}_t|\xi_t)\right]
\end{equation*}

One now easily verifies that the distribution $p(\overline{Z}_t|\xi_t)$ is governed by a backward recursive expression.
\begin{equation*}
\begin{aligned}
p(\overline{Z}_t|\xi_t) &= p(z_t|\xi_t) \int \tau_t(x_{t+1}|\xi_t) p(\overline{Z}_{t+1}|x_{t+1})\text{d}x_{t+1} \\
&= p(z_t|\xi_t) \expect{\tau_t(x_{t+1}|\xi_t)}\left[p(\overline{Z}_{t+1}|x_{t+1})\right]
\end{aligned}
\end{equation*}

Both the problem statements themselves, as well as the backward recursive calculation procedure, hint at a connection with the theory of dynamic programming. To illuminate the connection we may define the following negative $\log$-probabilities, where we used notation $l(\cdot)\equiv -\log p(\cdot)$. 
\begin{equation}
\begin{aligned}
Q^\star_{t}(\xi_t) &= l(\overline{Z}_{t}|\xi_t) = -\log \epsilon_t({z}_t|{\xi}_t)-\log \expect{\tau_t({x}'|{\xi}_t)}\left[\exp(-V^\star_{t+1}({x}'))\right] \\
V^\star_{t}({x}_t) &= l(\overline{Z}_{t}|{x}_{t}) = -\log \expect{\rho_t({u}_t|{x}_t)}\left[\exp(-{Q}^\star_t(\xi_t))\right]
\label{eq:Vstar}
\end{aligned}
\end{equation}
whereas the V-APCD is given by
\begin{equation}
\label{eq:Pstar}
\pi_t^\star({u}_t|{x}_t) = \rho_t({u}_t|{x}_t) \exp({V}^\star_t({x}_t)-{Q}^\star_t(\xi_t))
\end{equation}


\subsubsection{Information-theoretic argument}
Here we raise the question whether the V-APCD renders some objective function optimal? In particular we show that the V-APCD is governed by the M-projection of $p(\underline{\Xi}_T|\underline{Z}_T)$ onto the probability space spanned by $p(\underline{\Xi}_T;\underline{\pi}_T)$. Here $p(\underline{\Xi}_T;\underline{\pi}_T)$ is defined as the joint distribution obtained by administering \textit{some} probabilistic policy sequence $\underline{\pi}_T$ instead of the assumed probabilistic control model $\underline{\rho}_T(\underline{U}_T|\underline{X}_T)$\footnote{Note that therefore $p(\underline{\Xi}_T|\underline{Z}_T)$ is parametrized by the prior probabilistic control model $\underline{\rho}_T$. To keep notation light this subtlety was not included.}. Another way of looking at this is that we want to identify a probabilistic control model that induces a trajectory distribution that discriminates less from the posterior trajectory distribution. Casting the V-APCD as an optimization problem also allows to generalise the concept to multiple measurement sequences $\{\underline{Z}_T^n\}$. Assuming that the sequences are i.i.d. we propose to sum their contributions.

As such we consider following variational optimization problem that can be manipulated into $T-1$ separate subproblems.
%
\begin{equation}
\label{eq:1}
\begin{multlined}[.9\linewidth]
\arg\min_{\underline{\pi}_T\in\mathcal{P}} \sum\nolimits_n\mathbb{D}\left[p(\underline{\Xi}_T|\underline{Z}_T^n)\parallel p(\underline{\Xi}_T;\underline{\pi}_T)\right] \\
\begin{aligned}
&= \arg\max_{\underline{\pi}_T\in\mathcal{P}} \sum\nolimits_n\int p(\underline{\Xi}_T|\underline{Z}_T^n) \log \frac{p(\underline{\Xi}_T;\underline{\pi}_T)}{p(\underline{\Xi}_T|\underline{Z}_T^n)}\text{d}\underline{\Xi}_T \\
&= \arg\min_{\underline{\pi}_T\in\mathcal{P}}  \sum\nolimits_t \sum\nolimits_n\int  p(\underline{\Xi}_T|\underline{Z}_T^n) \log \pi_t(u_t|x_t) \text{d}\underline{\Xi}_T 
\end{aligned}
\end{multlined}
\end{equation}

The solution to this problem is governed by the V-APCD that we derived earlier for $N=1$ and by a mixture of individual V-APCDs for $N>1$. A derivation is provided in \ref{sec:derivation-of-refeqvapcd}. The mixture weights are determined by the individual smoothing distributions and can be calculated using traditional smoothing algorithms substituting the closed loop transition and emission models, $\tau_t' = \expect{\rho_t}[T_t]$ and $\epsilon_t' = \expect{\rho_t}[E_t]$. Thus the ensemble V-APCD determines how much each individual V-APCD contributes based on the probability that the would occupy the state, $x_t$, for each of the individual $N$ feature sequences $\underline{Z}_T^n$.
\begin{equation}
\label{eq:VAPCD}
\pi^\star_t(u_t|x_t) = \frac{\sum\nolimits_n p(\xi_t|\underline{Z}_T^n)}{\sum\nolimits_n p(x_t|\underline{Z}_T^n)} = \sum\nolimits_n\frac{p(x_t|\underline{Z}_T^n)}{\sum\nolimits_n p(x_t|\underline{Z}_T^n)} p(u_t|x_t,\overline{Z}_t^n)
\end{equation}

\subsection{Natural APCD}\label{sec:natural-apcd}
Reasoning from the information-theoretic motivation of the V-APCD above, we raise now the obvious question whether the reciprocal I-projection also generates an APCD? According to the information-theoretic interpretation of the I-projection it follows that therefore we minimize the inefficiency of assuming the prior probabilistic control model $\underline{\rho}_T$ whilst the true probabilistic control model is given by the posterior $\underline{\pi}_T$. This generates a sibling APCD. Provided our earlier discussion on terminology, we refer to the APCD derived here as the \textit{natural} APCD (N-APCD). We generalise the concept to multiple measurement sequences by superposing the individual contributions.

Thus we consider the variational optimization problem
\begin{equation}
\label{eq:5}
\min_{\underline{\pi}_T \in \mathcal{P}}  \sum\nolimits_n \mathbb{D}[p(\underline{\Xi}_T;\underline{\pi}_T)\parallel p(\underline{\Xi}_T|\underline{Z}_T^n)] 
\end{equation}
This problem can be recast as follows (\ref{sec:derivation-of-refeqsub})
\begin{equation}
\label{eq:sub}
\min_{\underline{\pi}_T \in \mathcal{P}} \expect{p(\underline{\Xi}_T;\underline{\pi}_T)} \left[- \tfrac{1}{N}\sum\nolimits_n\log p(\underline{Z}_T^n|\underline{\Xi}_T)+ \log \frac{\underline{\pi}_T(\underline{U}_T|\underline{X}_T))}{\underline{\rho}_T(\underline{U}_T|\underline{X}_T))} \right]
\end{equation}

The problem above exhibits an optimal substructure which permits application of the principle of dynamic programming. 
\newpage

In particular we can decompose the optimization problem as demonstrated below
\begin{equation*}
\begin{multlined}[.9\linewidth]
\min_{\underline{\pi}_t\in\mathcal{P}} \int  p(\underline{\Xi}_t,x_{t+1};\underline{\pi}_T)  \\\times \left(- \tfrac{1}{N}\sum\nolimits_n\log p(\overline{Z}_t^n|\overline{\Xi}_t)  + \log \frac{\underline{\pi}_t(\underline{U}_t|\underline{X}_t)}{\underline{\rho}_t(\underline{U}_t|\underline{X}_t)}\right)\text{d}\underline{\Xi}_t  V_{t+1}^\bullet(x_{t+1}) \text{d}x_{t+1}
\end{multlined}
\end{equation*}
Here we have defined the following value function
\begin{equation*}
V^\bullet_t(x_t) = \min_{\overline{\pi}_t \in \mathcal{P}} \expect{p(\overline{\Xi}_t|x_t,\overline{\pi}_t)} \left[- \tfrac{1}{N}\sum\nolimits_n\log p(\overline{Z}_t^n|\overline{\Xi}_t) + \log\frac{\overline{\pi}_t(\overline{U}_t|\overline{X}_t)}{\overline{\rho}_t(\overline{U}_t|\overline{X}_t)} \right]
\end{equation*}

Inspired by the structural connection with MDPs that was already hinted at in sec. \ref{sec:bayesian-approach}, one easily verifies that also $V_t^\bullet$ satisfies a backward recursive calculation procedure
\begin{equation*}
V^\bullet_t(x_t) = \min_{\pi_t \in \mathcal{P}} \expect{\pi_t(u_t|x_t)} \left[\log \frac{\pi_t(x_t|u_t)}{\rho_t(u_t|x_t)} + Q^\bullet_t(\xi_t) \right] 
\end{equation*}
where
\begin{equation}
Q^\bullet_t(\xi_t) = -\tfrac{1}{N}\sum\nolimits_n \log \epsilon_t(z_t^n|\xi_t) + \expect{\tau_t(x_{t+1}|\xi_t)}[V^\bullet_{t+1}(x_{t+1})]
\label{eq:Qbullet}
\end{equation}

Variational optimization (similar to \ref{sec:derivation-of-refeqvapcd}) of this final problem then yields an explicit expression for the value function $V_t^\bullet$ and thus the desired N-APCD, $\pi_t^\bullet$
\begin{equation}
\begin{aligned}
V^\bullet_t(x_t) &= - \log \expect{\rho_t(u_t|x_t)}\left[\exp(-Q^\bullet_t(\xi_t))\right] \\
\pi^\bullet_t(x_t|u_t) &= \rho_t(u_t|x_t) \exp(V^\bullet_t(x_t)-Q^\bullet_t(\xi_t))
\label{eq:Vbullet}
\end{aligned}
\end{equation}

\subsection{Some first observations}

We present here a number of observations regarding the methodologies set-out so far. 

\begin{enumerate}
	\item On account of the underlying probabilistic graph model, either APCD are governed by a \textit{Bayesian type} update rule. The probabilistic control model, ${\rho}_t$, acts as a prior probability which is updated to a posterior probabilistic control model, ${\pi}_t$, by the likelihood function, $\exp(-Q_t^*), * \in \{\star,\bullet\}$. 
	\item For $N=1$, the solution of the \textit{\textit{vanilla}} and \textit{\textit{natural}} problems appear equivalent. At least, so do the backwards recursive expressions for the associated policy distributions and value functions. The difference lies in the definition of the $Q$-functions. Comparing the expressions derived for $Q_t^\star$ (\ref{eq:Vstar}) and $Q_t^\bullet$ (\ref{eq:Qbullet}) reveals that the former additionally transforms the expectation in agreement with the probability-likelihood transformation (i.e. $l\equiv -\log p$) so that the apparent addition in likelihood space in fact amounts to a multiplication in probability space. In contrast, the \textit{\textit{natural}} approach is to carry out the computation in likelihood space. Further note that the difference is rendered irrelevant for deterministic dynamics in which case $ Q_t^\star \equiv Q_t^\bullet$ and therefore $ \pi_t^\star \equiv \pi_t^\bullet$ and so is $ V_t^\star \equiv V_t^\bullet$ 
	\item We can further analyse the difference in dynamics induced by both APCDs by considering the following decomposition of the posterior density, $p(\underline{\Xi}_T|\underline{Z}_T)$. We obtain
	\begin{equation*}
	p(\underline{\Xi}_T|\underline{Z}_T) = p(x_0|\underline{Z}_T)p(u_0|x_0,\underline{Z}_T)\prod\nolimits_t p(\xi_t|\xi_{t-1},\overline{Z}_t)
	\end{equation*}
	where
	\begin{equation*}
	p(\xi_t|\xi_{t-1},\overline{Z}_t) = p(x_t|{\xi}_{t-1},\overline{Z}_t) p(u_t|x_t,\overline{Z}_t)
	\end{equation*}
	\newpage
	This decomposition {thus serves as an a posteriori justification of the Bayesian argument in \ref{sec:bayesian-approach}. Unfortunately it also illustrates that the V-APCD} will only induce the true posterior trajectory distribution $p(\underline{\Xi}_T|\underline{Z}_T)$ if the system were governed by the \textit{informed} transition probability $p(x_t|\xi_{t-1},\overline{Z}_t)$. Clearly when the student practices the V-APCD, it cannot influence its own inherent dynamics so that we must substitute the natural transition probability rather than the informed transition probability. This means that the probability $p(\underline{\Xi}_T;\underline{\pi}_T^\star)$ differs from the probability $p(\underline{\Xi}_T|\underline{Z}_T)$. The reason is that the V-APCD aims to reconstruct the conditional distribution $p(u_t|x_t,\underline{Z}_T)$ and not the trajectory distribution itself. The V-APCD is naive in that sense. {A similar analysis is not possible for the N-APCD, though it is anticipated that the N-APCD tries to accommodate for the difference between the natural and informed transition probabilities. }
\end{enumerate}

\subsection{Linear-Gaussian Systems}\label{sec:hidden-gauss-markov-models}

{Here we specialize the APCDs to Linear-Gaussian (LG) systems
\begin{equation*}
\begin{aligned}
{x}_{t+1} &\sim T_t = \mathcal{N}(x_{t+1};\mathrm{F}_{\xi,t} \xi_{t} + {f}_{t},\mathrm{Q}_t)\\
{u}_{t} &\sim \rho_t =\mathcal{N}(u_t;\mathrm{K}_t {x}_t + {k}_t,\mathrm{S}_t)\\
{z}_{t} &\sim E_t =\mathcal{N}(z_t;\mathrm{G}_{\xi,t} \xi_{t} + {g}_t,\mathrm{R}_t) 
\end{aligned}
\end{equation*}
In this setting it is can be anticipated that the APCDs behave as affine Gaussian probabilities $\pi_t^* = \mathcal{N}({u}_t|\mathrm{K}_t^*{x}_t +{k}_t^*,\Sigma_t^*)$ and that both $Q_t^*$ and $V_t^*$ will be quadratic in their arguments.} 
{Computational details are given in \ref{sec:linear-gaussian-apcds}. Updates are given for $N=1$. Extension to (\ref{eq:Vbullet}) is trivial given (\ref{eq:Qbullet}). Extension to (\ref{eq:VAPCD}) requires calculating the smoothing distributions, $p(x_t|\underline{Z}_T^n)$ \cite{sarkka2013bayesian}.} 

With $*\in\{\star,\bullet\}$ throughout, the policy parameters are given
\begin{equation}
\label{eq:k}
\begin{aligned}
{k}_t^* &= \Sigma_t^*(\mathrm{S}_t^{-1}{k}_t - Q^*_{u,t}) \\
\mathrm{K}_t^* &= \Sigma_t^*(\mathrm{S}_t^{-1}\mathrm{K}_t - Q^*_{ux,t}) \\
\Sigma_t^* &= ( \mathrm{S}_t^{-1} + Q^*_{uu,t})^{-1} 
\end{aligned}
\end{equation}
where $Q_{\xi,t}^*$ and $Q_{\xi\xi,t}^*$ parametrise the quadratic model for $Q_t^*$. The parameters can be calculated recursively using the following expressions. Parameters $r_{\xi,t}$ and $r_{\xi\xi,t}$ relate to the quadratic model of the negative logarithm of the emission model. 
\begin{equation}
\label{eq:Q}
\begin{aligned} 
Q^*_{\xi,t} &= r_{\xi,t} + \mathrm{F}_{\xi,t}^\top(V_{xx,t+1}^{*,-1}+\mathbf{1}_\star(*)\cdot\mathrm{Q}_t)^{-1}\left(V_{xx,t+1}^{*,-1} V_{x,t}^\star+{f}_t\right) \\
Q_{\xi\xi,t}^* &= r_{\xi\xi,t} + \mathrm{F}_{\xi,t}^\top(V_{xx,t+1}^{*,-1}+\mathbf{1}_\star(*)\cdot\mathrm{Q}_t)^{-1}\mathrm{F}_{\xi,t} 
\end{aligned}
\end{equation}
Similarly $V_{\xi,t}^*$ and $V_{\xi\xi,t}^*$ are parameters from the quadratic model for $V_t^*$ and subject to the following recursions. 
\begin{equation}
\label{eq:V}
\begin{aligned} 
V_{x,t}^* &= Q_{x,t}^* + \mathrm{K}_t^{\top} \mathrm{S}_t^{-1} {k}_t - \mathrm{K}_t^{*,\top} \Sigma_t^{*,-1} {k}_t^* \\
V_{xx,t}^* &= Q_{xx,t}^* + \mathrm{K}_t^{\top} \mathrm{S}_t^{-1} \mathrm{K}_t - \mathrm{K}_t^{*,\top} \Sigma_t^{*,-1} {K}_t^* 
\end{aligned}
\end{equation}

{Practical procedures are given in Algorithm \ref{alg:LQ-V-APCD} and \ref{alg:LQ-N-APCD}. }

%

\section{Related work}

\subsection{Imitation Learning (from Observations)}
{
The goal of IL is to find a time-invariant policy function, $\pi:x\mapsto u$, (mapping states to actions), or, a sequence of time-varying policy functions, $\{\pi_t\}$, so that the closed-loop dynamics of the student produces behaviour similar to that of the teacher. 

\newpage

\begin{algorithm}[t]
	\begin{algorithmic}[1]
		\Require $\{\{\mathrm{F}_{\xi,t},f_t,\mathrm{Q}_t\}\}_t,\{\{\mathrm{K}_{t},k_t,\mathrm{S}_t\}\}_t,\{\{\mathrm{G}_{\xi,t},g_t,\mathrm{R}_t\}\}_t,\{z_{t}^n\}_{t,n}$
		\Ensure $\{\pi_t^\star\}_t$
		\For {$n\in\{1,2,\dots,N\}$}
		\State $p(\xi_t|\underline{Z}_T^n) = \mathcal{N}(\mu_{\xi,t},\Sigma_{\xi\xi,t}), \forall t \in {0,1,\dots,T}$ \cite{sarkka2013bayesian}
		\For {$t \in \{0,1,\dots,T-1\}$}
		\State $\mathcal{N}(u_t;k_t^{n,\star}+\mathrm{K}_t^{n,\star} x_t,\Sigma_t^{n,\star}) = p(u_t|x_t,\underline{Z}_T^n)$ \cite{petersen2008matrix}
		\State compute $\pi_t^\star$ according to (\ref{eq:VAPCD})
		\EndFor
		\EndFor
		\caption{LG-V-APCD}
		\label{alg:LQ-V-APCD}
	\end{algorithmic}
\end{algorithm}

\begin{algorithm}[t]
	\begin{algorithmic}[1]
		\Require $\{\{\mathrm{F}_{\xi,t},f_t,\mathrm{Q}_t\}\}_t,\{\{\mathrm{K}_{t},k_t,\mathrm{S}_t\}\}_t,\{\{\mathrm{G}_{\xi,t},g_t,\mathrm{R}_t\}\}_t,\{z_{t}^n\}_{t,n}$
		\Ensure $\{\pi_t^\bullet\}_t$
		\For {$t \in \{T,T-1,\dots,1\}$}
		\State update $\{V_{x,t}^{\bullet},V_{xx,t}^{\bullet}\}$ according to (\ref{eq:V})
		\State update $\{Q_{\xi,t-1}^{\bullet},Q_{\xi\xi,t-1}^{\bullet}\}$ according to (\ref{eq:Q}) 
		\State update $\{k_{t-1}^{\bullet},\mathrm{K}_{t-1}^{\bullet},\Sigma_{t-1}^{\bullet}\}$ according to (\ref{eq:k})
		\EndFor
		\caption{LG-N-APCD}
		\label{alg:LQ-N-APCD}
	\end{algorithmic}
\end{algorithm}

There are roughly two dominant approaches to face this problem, \textit{behavioural cloning} (BC) and \textit{inverse reinforcement learning} (IRL) \cite{torabi2019recent,osa2018algorithmic,pml2Book}. IRL methods model the behaviour of the student and teacher as a Markov Decision Process (MDP) and try to infer the cost/reward function that is used by the teacher to make policy in the belief that it is the most concise and portable representation of the task \cite{argall2009survey,ross2010efficient}. IRL offers insight as to why the teacher makes certain decisions, though the often time consuming policy (re)construction is delayed to post-processing.

BC is powerful in the sense that it requires only demonstration data to directly learn an imitation policy and does not require any further interaction between the agent and the environment. A natural approach to BC directly targets the mapping from states to actions through supervised learning. Learning of the policy $\pi$ reduces to solving the following problem. Recent advances in IL focus on the infinite horizon setting, $p(x,u;\mathcal{D})$ thus represents the stationary state-action data distribution.  
\begin{equation*}
J_{\text{BC}}[\pi;\mathcal{D}] = \expect{p(x,u;\mathcal{D})}[\log \pi(u|x)] 
\end{equation*}

\textit{Adversarial} IL (AIL) methods have shown great success in benchmarks for continuous control, especially in the low data regime \cite{ghasemipour2020divergence,pml2Book}. AIL directly aim to recover the policy similar to BC, yet are closely related to the MDP formulation of IRL. As it turns out many existing IL methods can be unified as IL by $f$-divergence minimization, so called \textit{distribution matching} \cite{ghasemipour2020divergence,kim2022lobsdice}. The reverse minimization is mode-seeking and preferable. 
\begin{equation*}
J_{\text{AIL}}[\pi;\mathcal{D}] = \left\lbrace
\begin{aligned}
&\mathbb{D}_f[p(x,u;\mathcal{D})||p(x,u;\pi)], && \text{forward} \\
&\mathbb{D}_f[p(x,u;\pi)||p(x,u;\mathcal{D})], && \text{reverse} 
\end{aligned}\right.
\end{equation*}

As stated in the introduction, IfO relaxes the requirement on action labels, and aims to imitate the expert’s behaviour only from the state observations. The amount of studies that focus on the IfO problem remains however limited \cite{torabi2019recent,torabi2018bc,Edwards2019im,kim2022lobsdice,zhu2020off,yang2019imitation,LiuLMS20}. A natural approach is to mimic BC by augmenting the state-only demonstrations with action labels. Roughly summarized inverse dynamic approaches invert consecutive states, $\{x_t,x_{t+1}\}$, into an action, $u_t$, and then progress along the direction of standard BC approaches by supervised learning \cite{ho2016generative,kim2013maximum,giusti2015machine,yang2019imitation}.  The reconstruction loss may range from simple least-squares regression to more complex losses such as e.g. inverse dynamics disagreement. 
More recently the concept of distribution matching has also been adopted. However, since the classical approach is no longer applicable, the distribution matching is recast in terms of the stationary state-transition distribution \cite{kim2022lobsdice,zhu2020off,yang2019imitation,LiuLMS20}. A unified view is given below. The hyperparameter $\alpha > 0$ balances between encouraging state-transition matching and preventing distribution shift from a set of imperfect demonstrations, $\mathcal{U}$.
\begin{equation*}
\begin{multlined}
J_{\text{IfO}}[\pi;\mathcal{D}] = \\ (1-\alpha)\mathbb{D}\left[p(x,x';\pi)||p(x,x';\mathcal{D})\right] + \alpha \mathbb{D}\left[p(x,u;\pi)||p(x,u;\mathcal{U})\right]
\end{multlined}
\end{equation*}

In light of the divergence minimization or distribution matching frameworks tailored to either IL and IfO reviewed above, it is interesting to revise the APCDs objectives from this work 
\begin{equation*}
J_{\text{IfPO}}[\pi;\mathcal{D},\rho] =\left\lbrace
\begin{aligned}
&\sum\nolimits_n\mathbb{D}[p(\underline{\Xi}_T|\underline{Z}^n_T;\underline{\rho}_T) || p(\underline{\Xi}_T;\underline{\pi}_T)], && \text{V-APCD} \\
&\sum\nolimits_n\mathbb{D}[p(\underline{\Xi}_T;\underline{\pi}_T)||p(\underline{\Xi}_T|\underline{Z}^n_T;\underline{\rho}_T)], && \text{N-APCD}
\end{aligned}\right.
\end{equation*}
 

}

\subsection{Connections with entropy regularized MDPs}\label{sec:connections-with-optimal-control}
Although we have deliberately not modelled the decision process of the student as an MDP, there have been indirect suggestions that either APCD solve some sort of Optimal Control problem. In order to make that intuition explicit, let us recall the theory of MDPs. An MDP is defined as follows
\begin{equation*}
\arg\min_{\overline{U}_0} \expect{p(\overline{X}_1|x_0,\overline{U}_0)}[{R}_T(\overline{\Xi}_T)]
\end{equation*}

The objective function is given by the cumulative cost $R_T(\underline{\Xi}_T) = \sum_t r_t(\xi_t)$. The solution is governed by a deterministic feedback policy sequence, $\{\pi^\blacktriangle_t\}_t$. Similar to the probabilistic APCDs, the sequence is governed by a backward recursion (\ref{eq:MDP}). Substituting $\expect{\pi_t}[Q^\bullet_t(\xi_t)] + \mathbb{D}[\pi_t||\rho_t]$ for $Q^\blacktriangle_t(\xi_t)$ and $l(z_t|\xi_t)$ for $r_t(\xi_t)$, we retrieve the optimization problem in (\ref{eq:Qbullet}). 
\begin{equation}
\label{eq:MDP}
\begin{aligned}
\pi^\blacktriangle_t(x_t) &= \arg\min_{u_t\in\mathcal{U}} Q^\blacktriangle_t(\xi_t) \\
V^\blacktriangle_t(x_t) &= \min_{u_t\in\mathcal{U}}Q^\blacktriangle_t(\xi_t) \\
Q^\blacktriangle_t(\xi_t) &= r_t(\xi_t) + \expect{p(x'|\xi_t)}[V^\blacktriangle_{t+1}(x')] 
\end{aligned}
\end{equation}

A risk-sensitive generalisation of the standard MDP has been developed where instead of minimizing the cumulative performance criteria $R_T(\Xi_T)$, the exponential of that objective, $\exp(-R_T(\Xi_T))$, is maximized \cite{jacobson1973optimal}. This choice puts more emphasis on the contribution of the tails of the trajectory distribution than is done when using the expected cumulative cost function. The solution of the risk-sensitive MDP is also given by a deterministic policy sequence, $\{\pi_t^\blacktriangledown\}_t$, governed by the recursive calculation in (\ref{eq:RSMDP}). Upon execution of the same set of substitutions that we introduced in the setting of standard MDPs, one verifies that we retrieve the same problem statement as in (\ref{eq:Vstar}).
\begin{equation}
\label{eq:RSMDP}
\begin{aligned}
\pi^\blacktriangledown_t(x_t) &= \arg\min_{u_t} Q^\blacktriangledown_t(\xi_t) \\
V^\blacktriangledown_t(x_t) &= \min_{u_t}Q^\blacktriangledown_t(\xi_t) \\
Q^\blacktriangledown_t(\xi_t) &= r_t(\xi_t) - \log \expect{p(x'|\xi_t)}[\exp(-V^\blacktriangledown_{t+1}(x'))] 
\end{aligned}
\end{equation}

These observations suggest that there is a very strong connection between the MDP frameworks detailed here and the APCDs derived in sections \ref{sec:vanilla-apcd} and \ref{sec:natural-apcd}, hence implying a strong connection between the BC approach that we originally set out for and IRL strategies that try to infer a cost model which is used consequently as input to solve an MDP. {For a single measurement sequence we conclude that the N-APCD behaves as an entropy regularized MPD, and the V-APCD behaves as an entropy regularized risk-sensitive MDP, using the conditional measurement negative $\log$-likelihood, $l(\underline{Z}_T|\underline{\Xi}_T)$, as cumulative cost. }Intuitively it is sensible to substitute $l(\underline{Z}_T|\underline{\Xi}_T)$ as a proxy for $R_T(\underline{\Xi}_T)$ representing how likely it is to have traversed some trajectory $\Xi_T$ given observation of the features $\underline{Z}_T$ similar to how eager we are to traverse that trajectory when we try to minimize $R_T(\underline{\Xi}_T)$. As a result of the entropy regularization we obtain expectation- rather than optimization operators and consequently we retrieve probabilistic policies rather than deterministic policies. {In conclusion we note that the connection with the V-APCD brakes down for multiple sequences given that the solution is then given as a mixture of individual V-APCDs. For the N-APCDs the connection is maintained with the cumulative cost averaged out over the multiple demonstrations.}

\section{Experiments}
In this section we document numerical experiments to validate the APCDs proposed in sections \ref{sec:vanilla-apcd} and \ref{sec:natural-apcd}. {With our numerical experiments we want to clear out which of the distributions is preferable, $p(\underline{\Xi}_T;\underline{\pi}_T^\star)$ or $p(\underline{\Xi}_T;\underline{\pi}_T^\bullet)$.} All experiments were implemented using \texttt{Matlab}. Each experiment was executed on a single 2.10GHz Intel Xeon Gold 6130 processor.

\begin{figure}[t]
	\centering
	\includegraphics[width=\columnwidth]{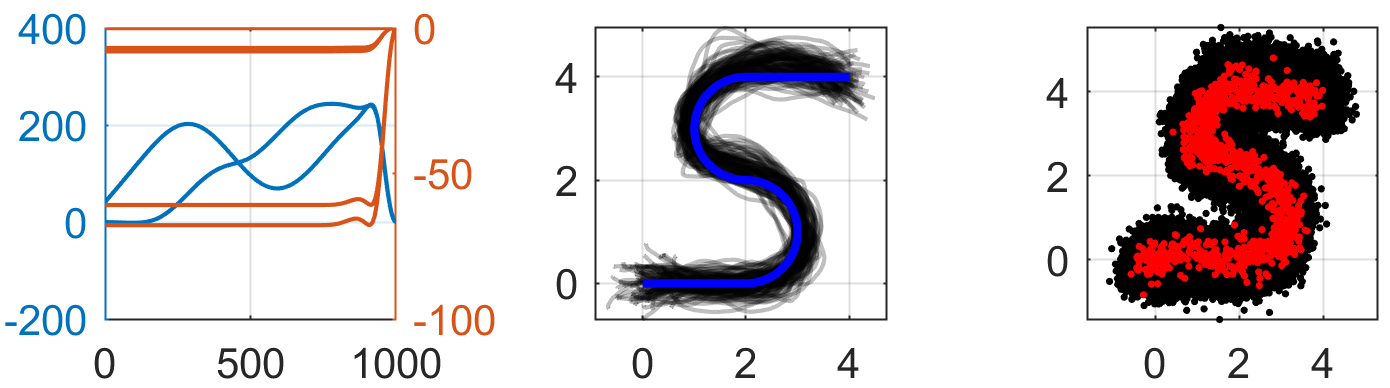}
	\caption{Illustrative linear problem. We want to reconstruct the behavioural features of the force control feedback policy from position measurements. \textbf{Left}: feedforward terms (blue) and feedback gains (red). \textbf{Middle}: $10^2$ tracking experiments (black) about a reference path (blue). \textbf{Right}: $10^2$ measurement sequences (black) with highlighted first measurement sequence (red).}
	\label{fig:sim}
\end{figure}

\subsection{Problem definition} We consider a force controlled planar mass with Brownian input noise, see Fig. \ref{fig:sim}. The covariance of the input noise is spawned cascading the \texttt{rand} and \texttt{sprandsym} command generating correlated white noise. To realise anisotropy, the random seed is multiplied with $\mathrm{diag}([10,10^2])$. The system is discretised using a sample period $\Delta t = 2\cdot 10^{-3} \si{\second}$. We define the teacher as a path tracking problem over the horizon $T = 2 \si{\second}$. The teacher's policy is given  by a Linear Quadratic Exponential Regulator (LQER) \cite{jacobson1973optimal} minimizing the objective defined below. {Here $p_t$, $p_t^*$ and $v_t$ define the position, desired position and velocity of the particle respectively. 
We set $\mathrm{W}_p = 10^4 \cdot\mathrm{I}$, $\mathrm{W}_p = \mathrm{W}_u = \mathrm{I}$ and $\lambda = 10^{-4}$.} It is well known that the solution is given by a time dependent linear policy, i.e. $u_t = k_t^\square + \mathrm{K}_t^\square x_t$. When simulating the system we use exact state observations though the system is tracked using position measurements, i.e. $z_t = p_t$, for post-processing. The covariance of the measurement noise was determined using the same procedure as the input noise, though here the random seed was multiplied with $10^{-1}$ in all dimensions. Finally we assume that the system is initialised with zero mean white noise with $\sigma_0^2 = 10^{-1}$.
\begin{equation*}
\min \mathbb{E} \left[\exp\left(\frac{\lambda}{2}\sum\nolimits_{t=0}^{T} \|p_t - p_t^*\|^2_{\mathrm{W}_p} + \|v_t\|^2_{\mathrm{W}_v} + \|u_t\|^2_{\mathrm{W}_u} \right)\right]
\end{equation*}

\subsection{Results}\label{sec:results}

We verify the capacity of the V-APCD and N-APCD to reconstruct the underlying policy $\{\underline{k}_T^\square,\underline{\mathrm{K}}_T^\square\}$ using the APCD mean as a proxy for the LQER. Thus we interpret the APCD covariance as a measure for our epistemic uncertainty about the APCD mean. For the N-APCD this results into a linear feedback policy. For the V-APCD this results into a Gaussian mixture of $N$ linear feedback policies. The policy prior, $\underline{\rho}_T$, is characterised as a linear Gaussian policy with zero mean and uncorrelated covariance with magnitude $\sigma^2$. Because we are interested in behavioural features of dynamics induced by the learned student's policy, rather than in a perfect reconstruction of the teacher's policy we do not quantify the reconstruction of the LQER itself but compare the performance of the APCDs with respect to the control objective defined above. In our experiments we vary two hyperparameters, in particular the magnitude $\sigma^2$ and the number of measurement sequences $N$ picked randomly from half of $M = 10^2$ individual experiments. The APCDs are then validated on the same $10^2$ experiments using the same in- and output noise. To counteract the influence of the specific $N$ from half the $M$ sequences on the reconstruction, we verify $P$ unique but random combinations so that the probability of never having included a specific sequence is less than $1\%$\footnote{The number of unique combinations when picking $N$ from $M$ is $B = \binom{M}{N}$. The number of unique combinations that does not include a specific experiment is $A = \binom{M-1}{N}$. The probability of picking a combination that does not include a specific experiment on the $(p-1)$\textsuperscript{th} try equals $q(p) = \frac{A-p}{B-p}$. The probability of picking $P$ unique combinations that do not include a specific sequence equals $f(P) = \prod_{p=1}^{P} q(p)$. As such we can compute $P$ so that $f(P) \leq \overline{f}$ for given $N$.}. See the inline figure.
\begin{wrapfigure}{R}{0pt}
	\centering
	\includegraphics[width=.35\columnwidth]{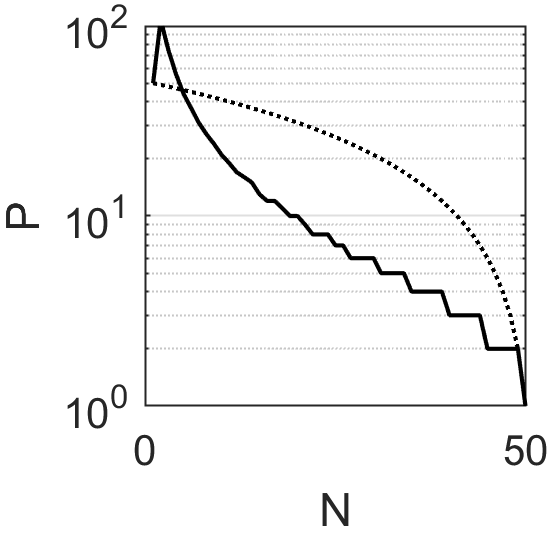}
\end{wrapfigure}

\begin{figure}[t]
	\centering
	\begin{subfigure}{.495\columnwidth}
		\centering
		\includegraphics[width=\columnwidth]{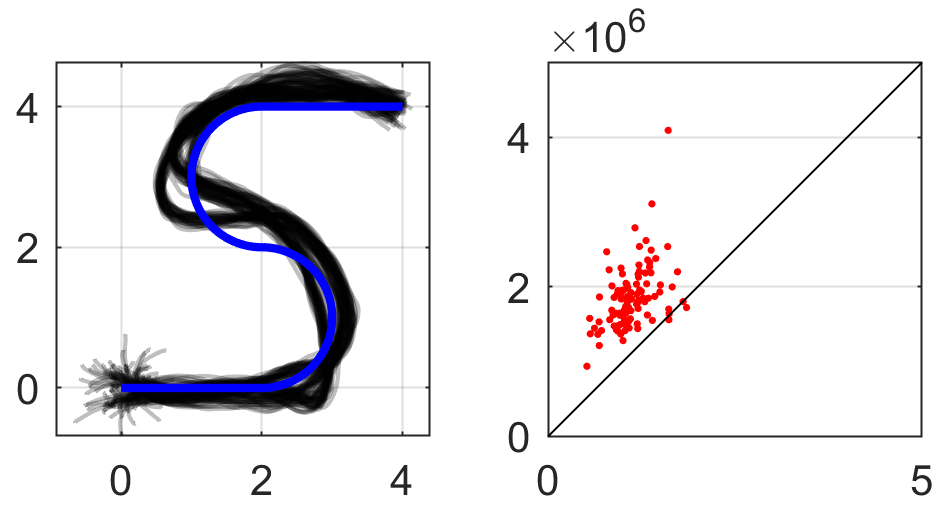}
		\caption{V-APCD}
		\label{fig:sim1}
	\end{subfigure}
	\begin{subfigure}{.495\columnwidth}
		\centering
		\includegraphics[width=\columnwidth]{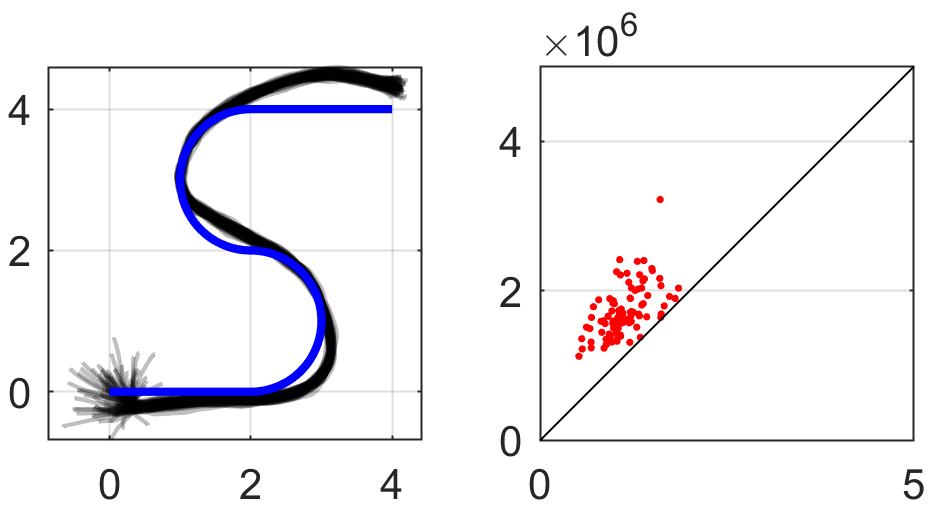}
		\caption{N-APCD}
		\label{fig:sim2}
	\end{subfigure}
	\caption{Comparison of the V- and N-APCD for $\sigma^2=10^4$ and $N=3$. The figures on the right compares the cost obtained with the learned student's policy (vertical) and the cost obtained with the teacher's policy (horizontal) when repeating the experiment with the same noise values. For experiments on the diagonal the same cost was obtained. For experiments above the diagonal, the teacher outperforms the student.}
	\label{fig:sim12}
	\vspace{12pt}
	\centering
	\begin{subfigure}{.495\columnwidth}
		\centering
		\includegraphics[width=\columnwidth]{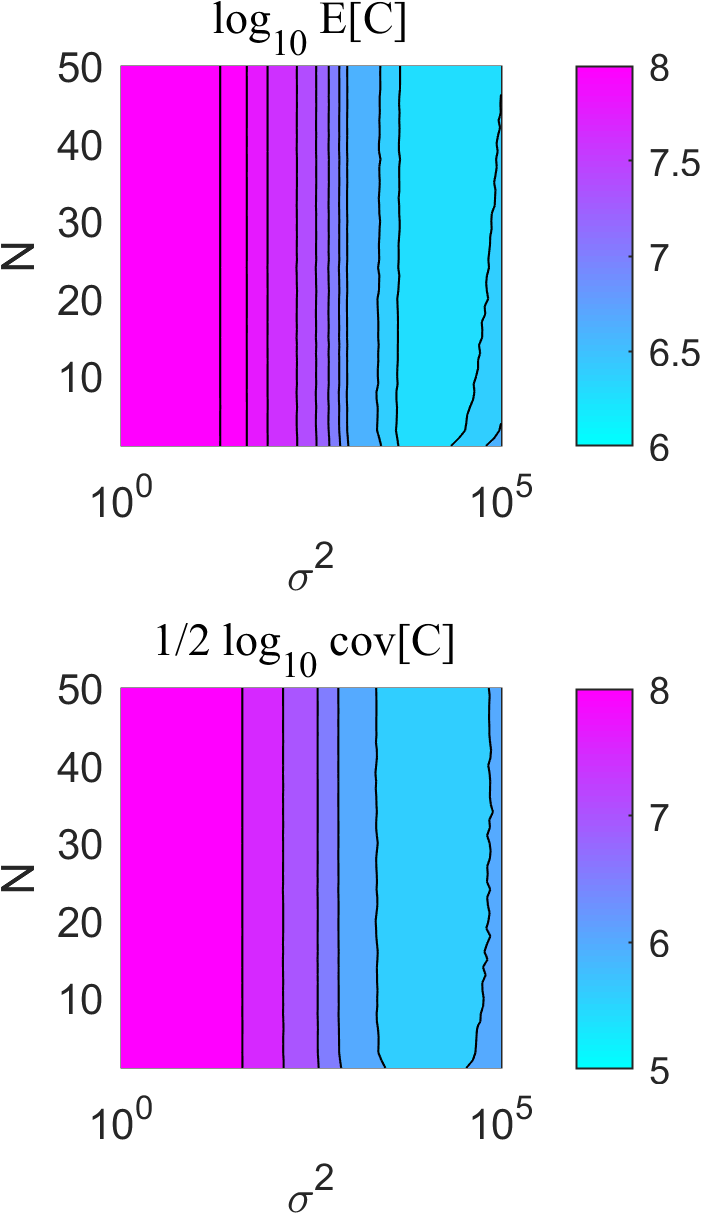}
		\caption{V-APCD}
		\label{fig:res_apcd1}
	\end{subfigure}
	\begin{subfigure}{.495\columnwidth}
		\centering
		\includegraphics[width=\columnwidth]{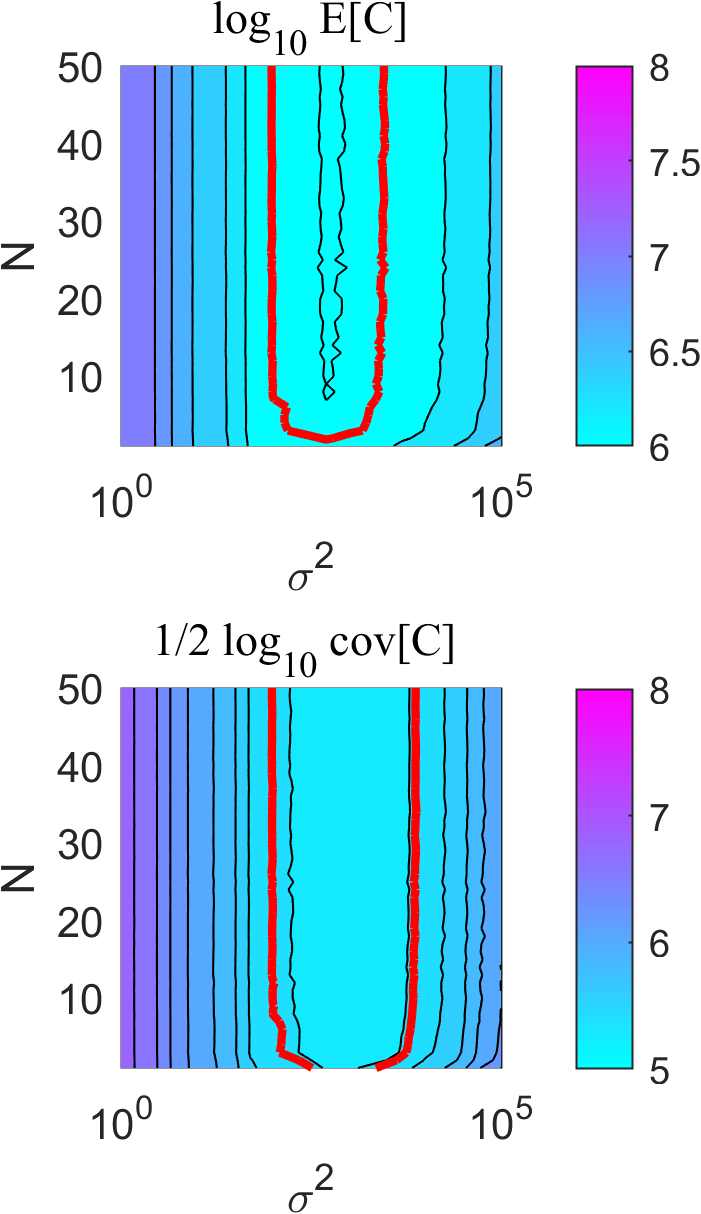}
		\caption{N-APCD}
		\label{fig:res_apcd2}
	\end{subfigure}
	\caption{Comparison of the V- and N-APCD performance as measured by the original control objective for varying $\sigma^2$ and $N$ according to the principles described in section \ref{sec:results}. The red curves indicate the performance of the LQER for $M=100$. Note that the horizontal scale is logarithmic.}
	\label{fig:res}
\end{figure}

Fig. \ref{fig:sim12} visualizes the performance of both the V- and N-APCD respectively for $\sigma^2 = 10^4$ and $N=3$. It is interesting to note that either APCD is capable of reconstructing the behavioural features of the LQER to visual satisfaction. One can verify that the V-APCD acts as a combination of 3 individual policies where the acting policy is determined by the measurement sequences that best explains the current state according to $p(x_t|\underline{Z}_T^n)$. On the contrary the N-APCD averages out the contribution of each measurement sequence in likelihood space. As can be seen, for smaller $N$ this results into a slight misalignment of the reconstructed desired behaviour and the true reference path. Fig. \ref{fig:res} documents results for varying $\sigma^2$ and $N$. Depending on $N$ the experiment was repeated $P$ times with a unique combination of sequences. Overall the N-APCDs can be seen to outperform the V-APCDs with the N-APCDs obtaining similar performance to the LQER.

\section{Conclusion}
In this work we {formalized the problem of Imitation from Partial Observations} and discussed two novel strategies to treat the Imitation from Observations problem from a single or multiple partial observations sequences. We have set-up a Behavioural Cloning approach based on the probabilistic graph model in Fig. \ref{fig:CHMM}. By reformulating the problem as an information-theoretic Moment projection, the strategy could be generalised to multiple observations sequences. Treatment of the reciprocal Information projection yielded a sibling solution to the imitation problem. {Comparison with recent work on distribution matching tailored to Imitation Learning (from Observations) illustrates that our work can be classified as a distribution matching method for Imitation from Partial Observations.}

Further investigation illustrated that both policy learning methods can also be interpreted as a form of Inverse Reinforcement Learning using a specific choice for the cost model and using entropy regularization to bias the inferred policy on some prior policy probability. {Provided that the policies can also be seen as specific instances of entropy regularized Markov Decision Processes, it should be possible to add auxiliary features to the student's policy by extending the cost model with additional terms that reflect specific behaviour.}

Specialization of our results to Linear-Gaussian models provided an explicit backward recursive calculation procedure to infer the desired policy distributions which was verified empirically. {Preliminary sufficient conditions for which these procedures will yield feasible solutions were given leaning on classical estimation and control theory. A more detailed analysis is still required to determine necessary conditions.} 

In future work it would also be interesting to extend the treatment to nonlinear probabilistic state-space models as well and to further explore the theoretical connection with the generic representation learning problem where neither the transition and/or the emission model are known. 

\section*{Acknowledgements} The authors wish to acknowledge financial support from the Research Foundation -- Flanders (FWO), grant no. S007723N.

\bibliographystyle{elsarticle-num}
\bibliography{references}

\begin{thebibliography}{10}
\expandafter\ifx\csname url\endcsname\relax
  \def\url#1{\texttt{#1}}\fi
\expandafter\ifx\csname urlprefix\endcsname\relax\def\urlprefix{URL }\fi
\expandafter\ifx\csname href\endcsname\relax
  \def\href#1#2{#2} \def\path#1{#1}\fi

\bibitem{argall2009survey}
B.~D. Argall, S.~Chernova, M.~Veloso, B.~Browning, A survey of robot learning
  from demonstration, Robotics and autonomous systems 57~(5) (2009) 469--483.

\bibitem{osa2018algorithmic}
T.~Osa, J.~Pajarinen, G.~Neumann, J.~A. Bagnell, P.~Abbeel, J.~Peters, et~al.,
  An algorithmic perspective on imitation learning, Foundations and
  Trends{\textregistered} in Robotics 7~(1-2) (2018) 1--179.

\bibitem{schaal1996learning}
S.~Schaal, Learning from demonstration, Advances in neural information
  processing systems 9.

\bibitem{pan2020imitation}
Y.~Pan, C.-A. Cheng, K.~Saigol, K.~Lee, X.~Yan, E.~A. Theodorou, B.~Boots,
  Imitation learning for agile autonomous driving, The International Journal of
  Robotics Research 39~(2-3) (2020) 286--302.

\bibitem{zhang2018deep}
T.~Zhang, Z.~McCarthy, O.~Jow, D.~Lee, X.~Chen, K.~Goldberg, P.~Abbeel, Deep
  imitation learning for complex manipulation tasks from virtual reality
  teleoperation, in: 2018 IEEE International Conference on Robotics and
  Automation (ICRA), IEEE, 2018, pp. 5628--5635.

\bibitem{jiang2021manipulator}
D.~Jiang, G.~Li, Y.~Sun, J.~Hu, J.~Yun, Y.~Liu, Manipulator grabbing position
  detection with information fusion of color image and depth image using deep
  learning, Journal of Ambient Intelligence and Humanized Computing 12~(12)
  (2021) 10809--10822.

\bibitem{zhang2019effective}
X.~Zhang, J.~Liu, J.~Feng, Y.~Liu, Z.~Ju, Effective capture of nongraspable
  objects for space robots using geometric cage pairs, IEEE/ASME Transactions
  on Mechatronics 25~(1) (2019) 95--107.

\bibitem{sermanet2018time}
P.~Sermanet, C.~Lynch, Y.~Chebotar, J.~Hsu, E.~Jang, S.~Schaal, S.~Levine,
  G.~Brain, Time-contrastive networks: Self-supervised learning from video, in:
  2018 IEEE international conference on robotics and automation (ICRA), IEEE,
  2018, pp. 1134--1141.

\bibitem{torabi2019recent}
F.~Torabi, G.~Warnell, P.~Stone,
  \href{https://doi.org/10.24963/ijcai.2019/882}{Recent advances in imitation
  learning from observation}, in: Proceedings of the Twenty-Eighth
  International Joint Conference on Artificial Intelligence, {IJCAI-19},
  International Joint Conferences on Artificial Intelligence Organization,
  2019, pp. 6325--6331.
\newblock \href {http://dx.doi.org/10.24963/ijcai.2019/882}
  {\path{doi:10.24963/ijcai.2019/882}}.
\newline\urlprefix\url{https://doi.org/10.24963/ijcai.2019/882}

\bibitem{sharma2019third}
P.~Sharma, D.~Pathak, A.~Gupta, Third-person visual imitation learning via
  decoupled hierarchical controller, Advances in Neural Information Processing
  Systems 32.

\bibitem{torabi2018bc}
F.~Torabi, G.~Warnell, P.~Stone, Behavioral cloning from observation, in:
  Proceedings of the 27th International Joint Conference on Artificial
  Intelligence, IJCAI'18, AAAI Press, 2018, p. 4950–4957.

\bibitem{liu2018imitation}
Y.~Liu, A.~Gupta, P.~Abbeel, S.~Levine, Imitation from observation: Learning to
  imitate behaviors from raw video via context translation, in: 2018 IEEE
  International Conference on Robotics and Automation (ICRA), IEEE, 2018, pp.
  1118--1125.

\bibitem{sun2019provably}
W.~Sun, A.~Vemula, B.~Boots, D.~Bagnell, Provably efficient imitation learning
  from observation alone, in: International conference on machine learning,
  PMLR, 2019, pp. 6036--6045.

\bibitem{pierrot2021}
T.~Pierrot, N.~Perrin-Gilbert, O.~Sigaud, First-order and second-order variants
  of the gradient descent in a unified framework, in: I.~Farka{\v{s}},
  P.~Masulli, S.~Otte, S.~Wermter (Eds.), Artificial Neural Networks and
  Machine Learning -- ICANN 2021, Springer International Publishing, Cham,
  2021, pp. 197--208.

\bibitem{cover2006elements}
T.~M. Cover, J.~A. Thomas, Elements of Information Theory (Wiley Series in
  Telecommunications and Signal Processing), Wiley-Interscience, USA, 2006.

\bibitem{bishop2006pattern}
C.~M. Bishop, N.~M. Nasrabadi, Pattern recognition and machine learning,
  Vol.~4, Springer, 2006.

\bibitem{pml1Book}
K.~P. Murphy, Probabilistic Machine Learning: An introduction, MIT Press, 2022.

\bibitem{sarkka2013bayesian}
S.~S{\"a}rkk{\"a}, Bayesian filtering and smoothing, no.~3, Cambridge
  University Press, 2013.

\bibitem{petersen2008matrix}
K.~B. Petersen, M.~S. Pedersen, et~al., The matrix cookbook, Technical
  University of Denmark 7~(15) (2008) 510.

\bibitem{pml2Book}
K.~P. Murphy, Probabilistic Machine Learning: Advanced Topics, MIT Press, 2023.

\bibitem{ross2010efficient}
S.~Ross, D.~Bagnell, Efficient reductions for imitation learning, in:
  Proceedings of the thirteenth international conference on artificial
  intelligence and statistics, JMLR Workshop and Conference Proceedings, 2010,
  pp. 661--668.

\bibitem{ghasemipour2020divergence}
S.~K.~S. Ghasemipour, R.~Zemel, S.~Gu, A divergence minimization perspective on
  imitation learning methods, in: Conference on Robot Learning, PMLR, 2020, pp.
  1259--1277.

\bibitem{kim2022lobsdice}
G.-H. Kim, J.~Lee, Y.~Jang, H.~Yang, K.-E. Kim, Lobsdice: Offline imitation
  learning from observation via stationary distribution correction estimation,
  arXiv preprint arXiv:2202.13536.

\bibitem{Edwards2019im}
A.~Edwards, H.~Sahni, Y.~Schroecker, C.~Isbell, Imitating latent policies from
  observation, in: K.~Chaudhuri, R.~Salakhutdinov (Eds.), Proceedings of the
  36th International Conference on Machine Learning, Vol.~97 of Proceedings of
  Machine Learning Research, PMLR, 2019, pp. 1755--1763.

\bibitem{zhu2020off}
Z.~Zhu, K.~Lin, B.~Dai, J.~Zhou, Off-policy imitation learning from
  observations, Advances in Neural Information Processing Systems 33 (2020)
  12402--12413.

\bibitem{yang2019imitation}
C.~Yang, X.~Ma, W.~Huang, F.~Sun, H.~Liu, J.~Huang, C.~Gan, Imitation learning
  from observations by minimizing inverse dynamics disagreement, Advances in
  neural information processing systems 32.

\bibitem{LiuLMS20}
F.~Liu, Z.~Ling, T.~Mu, H.~Su,
  \href{https://openreview.net/forum?id=rylrdxHFDr}{State alignment-based
  imitation learning}, in: 8th International Conference on Learning
  Representations, ICLR 2020, Addis Ababa, Ethiopia, April 26-30, 2020,
  OpenReview.net, 2020.
\newline\urlprefix\url{https://openreview.net/forum?id=rylrdxHFDr}

\bibitem{ho2016generative}
J.~Ho, S.~Ermon, Generative adversarial imitation learning, Advances in neural
  information processing systems 29.

\bibitem{kim2013maximum}
B.~Kim, J.~Pineau, Maximum mean discrepancy imitation learning., in: Robotics:
  Science and systems, 2013.

\bibitem{giusti2015machine}
A.~Giusti, J.~Guzzi, D.~C. Cire{\c{s}}an, F.-L. He, J.~P. Rodr{\'\i}guez,
  F.~Fontana, M.~Faessler, C.~Forster, J.~Schmidhuber, G.~Di~Caro, et~al., A
  machine learning approach to visual perception of forest trails for mobile
  robots, IEEE Robotics and Automation Letters 1~(2) (2015) 661--667.

\bibitem{jacobson1973optimal}
D.~Jacobson, Optimal stochastic linear systems with exponential performance
  criteria and their relation to deterministic differential games, IEEE
  Transactions on Automatic control 18~(2) (1973) 124--131.

\bibitem{barfoot2017state}
T.~Barfoot, State estimation for robotics, Cambridge University Press, 2017.

\bibitem{lefebvre2022}
T.~Lefebvre, On optimal control and expectation-maximisation: Theory and an
  outlook towards algorithms (2022).
\newblock \href {http://dx.doi.org/10.48550/ARXIV.2205.03279}
  {\path{doi:10.48550/ARXIV.2205.03279}}.

\end{thebibliography}

\newpage
\appendix

\section{Derivation of (\ref{eq:VAPCD})}\label{sec:derivation-of-refeqvapcd}
We retake from equation (\ref{eq:1})
\begin{equation*}
\min_{\underline{\pi}_T\in\mathcal{P}}  \sum\nolimits_t \sum\nolimits_n\int  p(\underline{\Xi}_T|\underline{Z}_T^n) \log \pi_t(u_t|x_t) \text{d}\underline{\Xi}_T 
\end{equation*}

Clearly each term in the first summation can be treated separately. We obtain the following $T$ independent  subproblems 
\begin{equation*}
\min_{{\pi}_t\in\mathcal{P}} \sum\nolimits_n\int  p(\underline{\Xi}_T|\underline{Z}_T^n) \log \pi_t(u_t|x_t) \text{d}\underline{\Xi}_T 
\end{equation*}

Then because the optimization variable $\pi_t$ depends only on the terms $\xi_t$ we can marginalize over the leading and trailing sequences $\underline{\Xi}_{t-1}$ and $\overline{\Xi}_{t+1}$. This yields the integrand
\begin{equation*}
\int \sum\nolimits_n  p({\xi}_t|\underline{Z}_T^n) \log \pi_t(u_t|x_t) \text{d}{\xi}_t 
\end{equation*}

Finally because $\pi_t\in\mathcal{P}$ we introduce the Lagrangian multiplier $\lambda_t$ and consider the Lagrangian
\begin{equation*}
\mathcal{L}_t[\pi_t,\lambda_t] = \int \sum\nolimits_n p({\xi}_t|\underline{Z}_T^n) \log \pi_t(u_t|x_t) \text{d}{\xi}_t - \lambda_t \int \pi_t(u_t|x_t)\text{d}u_t
\end{equation*}

On account of the calculus of variations, we have that the derivative of the integrand should equal $0$.
\begin{equation*}
\pi_t(u_t|x_t) = \frac{1}{\lambda_t}\sum\nolimits_n p({\xi}_t|\underline{Z}_T^n), ~ \lambda_t = \int \sum\nolimits_n p({\xi}_t|\underline{Z}_T^n) \text{d}u_t
\end{equation*}

\section{Derivation of (\ref{eq:sub})}\label{sec:derivation-of-refeqsub}

We retake from equation (\ref{eq:5}) where we write out the expression for the divergence explicitly
\begin{equation*}
\min_{\underline{\pi}_T \in \mathcal{P}}  \sum\nolimits_n \int p(\underline{\Xi}_T;\underline{\pi}_T) \log \frac{p(\underline{\Xi}_T;\underline{\pi}_T)}{p(\underline{\Xi}_T|\underline{Z}_T^n)}\text{d}\underline{\Xi}_T
\end{equation*}
where we may further specify the denominator
\begin{equation*}
\begin{aligned}
p(\underline{\Xi}_T|\underline{Z}_T^n) &= \tfrac{1}{p(\underline{Z}_T^n)} p(\underline{\Xi}_T,\underline{Z}_T^n)= \tfrac{1}{p(\underline{Z}_T^n)} p(\underline{Z}_T^n|\underline{\Xi}_T)p(\underline{\Xi}_T;\underline{\rho}_T)
\end{aligned}
\end{equation*}
Then we substitute this expression back into the original optimization problem and rearrange terms. We retrieve
\begin{equation*}
\min_{\underline{\pi}_T \in \mathcal{P}}  \sum\nolimits_n \int p(\underline{\Xi}_T;\underline{\pi}_T) \log \frac{p(\underline{\Xi}_T;\underline{\pi}_T)}{p(\underline{\Xi}_T,\underline{Z}_T^n)}\text{d}\underline{\Xi}_T + \sum\nolimits_n \log p(\underline{Z}_T^n)
\end{equation*}

Since the trailing term does not depend on $\underline{\pi}_T$ we can neglect it further. A final rearrangement of terms yields (\ref{eq:sub})
\begin{equation*}
\min_{\underline{\pi}_T \in \mathcal{P}} \int p(\underline{\Xi}_T;\underline{\pi}_T) \sum\nolimits_n\left( -\log p(\underline{Z}_T^n|\underline{\Xi}_T) +\log \frac{p(\underline{\Xi}_T;\underline{\pi}_T)}{p(\underline{\Xi}_T;\underline{\rho}_T)}\right)\text{d}\underline{\Xi}_T 
\end{equation*}

\section{Linear-Gaussian APCDs}\label{sec:linear-gaussian-apcds}

\subsection{Derivation of (\ref{eq:k}), (\ref{eq:Q}) and (\ref{eq:V})}
We restart from the conditions described in the beginning of section \ref{sec:hidden-gauss-markov-models}. Further we introduce the symbol, $\times$, to refer to constants that have no relevant effect on the calculations. 

First we assumed that both $Q_t^*$ and $V_t^*$ are quadratic. 
\begin{equation*}
\begin{aligned}
V^*_t({x}_t) &= \tfrac{1}{2}\begin{bmatrix}
1 \\ {x}_{t} 
\end{bmatrix}^\top \begin{bmatrix}
\times & V_{x,t}^{*,\top} \\
V_{x,t}^{*} & V_{xx,t}^*
\end{bmatrix}\begin{bmatrix}
1 \\ {x}_{t} 
\end{bmatrix}\\
Q^*_t({\xi}_t) &= \tfrac{1}{2}\begin{bmatrix}
1 \\ {\xi}_{t} 
\end{bmatrix}^\top \begin{bmatrix}
\times & Q_{\xi,t}^{*,\top}  \\
Q_{\xi,t}^{*} & Q_{\xi\xi,t}^* \\
\end{bmatrix}\begin{bmatrix}
1 \\ {\xi}_{t}
\end{bmatrix}\\
\end{aligned}
\end{equation*}
\newpage

Assuming then that we have gained access to the values $Q_{\xi,t}^*$ and $Q_{\xi\xi,t}^*$ we can derive expressions that hold irrespective whether we consider the \textit{vanilla} or \textit{natural} APCD. Since $\pi_t^*(u_t|x_t) \propto \rho(u_t|x_t) \exp(-Q^*_t(\xi_t))$ one verifies the policy (\ref{eq:k}). Second, since $\exp(-V^*_t(x_t)) \pi_t^*(u_t|x_t) = \rho(u_t|x_t) \exp(-Q^*_t(\xi_t))$ for any $u_t$ including $0$, one easily verifies the value update in (\ref{eq:V}).
Only the update for the $Q$-function differs between the \textit{vanilla} and \textit{natural} APCDs. We can derive expression based on the definitions in (\ref{eq:Vstar}) and (\ref{eq:Qbullet}) respectively. Therefore we further introduce a quadratic expression for $l({z}_t|{\xi}_t)$. Note that the parameters $r_{\xi,t}$ and $r_{\xi\xi,t}$ depend on the model $\mathrm{G}_{\xi,t}$, $g_t$ and ${R}_t$.
\begin{equation*}
l({z}_t|{\xi}_t) = r_t(\xi_t) = \tfrac{1}{2}\begin{bmatrix}
1 \\ {\xi}_{t} 
\end{bmatrix}^\top \begin{bmatrix}
\times & r_{\xi,t}^{\top} \\
r_{\xi,t} & r_{\xi\xi,t}
\end{bmatrix}\begin{bmatrix}
1 \\ {\xi}_{t} 
\end{bmatrix}
\end{equation*}

\paragraph{V-APCD} For the \textit{vanilla} $Q$-function we find 
\begin{equation*}
\begin{aligned} 
Q^\star_{\xi,t} &= r_{\xi,t} + \mathrm{F}_{\xi,t}^\top(V_{xx,t+1}^{\star,-1}+\mathrm{Q}_t)^{-1}\left(V_{xx,t+1}^{\star,-1} V_{x,t}^\star+{f}_t\right) \\
Q_{\xi\xi,t}^\star &= r_{\xi\xi,t} + \mathrm{F}_{\xi,t}^\top(V_{xx,t+1}^{\star,-1}+\mathrm{Q}_t)^{-1}\mathrm{F}_{\xi,t} 
\end{aligned}
\end{equation*}
For a single measurement the V-APCD is a linear-Gaussian policy. For multiple measurements its a Gaussian mixture model.

\paragraph{N-APCD} For the \textit{natural} $Q$-function we have that
\begin{equation*}
\begin{aligned} 
Q^\bullet_{\xi,t} &= r_{\xi,t} + \mathrm{F}_{\xi,t}^\top \left(V_{xx,t+1}^\bullet{f}_t+ V_{x,t}^\bullet\right) \\
Q_{\xi\xi,t}^\bullet &= r_{\xi\xi,t} + \mathrm{F}_{\xi,t}^\top V_{xx,t+1}^\bullet \mathrm{F}_{\xi,t} 
\end{aligned}
\end{equation*}
This expression holds both for single as well as multiple measurements. In the latter case $r_t$ is given by the average (\ref{eq:Qbullet}). In either case the N-APCD is given by a linear-Gaussian policy.

{
\subsection{Existence of solution}
In conclusion we give here a sufficient condition for the existence of the solution of the problem (\ref{eq:VAPCD}) for LG dynamic systems. It is shown that $\pi_t^\star$ can be computed from the individual smoothing distribution $p(\xi_t|\underline{Z}_T^n)$. In the linear-Gaussian setting these distributions exist if the system is observable \cite{barfoot2017state}. Consequently the APCD exists if the auxiliary dynamics system $\xi_t$ is observable. The auxiliary dynamics are governed by 
\begin{equation*}
\xi_{t+1} \sim \mathcal{N}(\xi_{t+1};\mathrm{A}_t \xi_t + a_t,\mathrm{P}_t) 
\end{equation*}
where 
\begin{equation*}
\begin{aligned}
\mathrm{A}_t &= \begin{bmatrix}
\mathrm{F}_{\xi,t} \\
\mathrm{K}_{t+1} \mathrm{F}_{\xi,t}
\end{bmatrix} \\ a_t &= \begin{bmatrix}
f_{\xi,t} \\
\mathrm{K}_{t+1} f_{\xi,t} + k_{t+1}
\end{bmatrix}\\
\mathrm{P}_t &= \begin{bmatrix}
\mathrm{Q}_t & \mathrm{Q}_t\mathrm{K}_{t+1}^\top \\
\mathrm{K}_{t+1} \mathrm{Q}_t & \mathrm{R}_t + \mathrm{K}_{t+1} \mathrm{Q}_t \mathrm{K}_{t+1}^\top 
\end{bmatrix}
\end{aligned}
\end{equation*}
If $\{\mathrm{A}_t,\mathrm{G}_t\}$ is observable, a solution exists for (\ref{eq:VAPCD}).

A similar analysis of (\ref{eq:Qbullet}) for Linear-Gaussian systems is impossible leaning on classical estimation theory. In this case the duality between the entropy regularized MDP and the V-APCD can be exploited. It was shown that if the underlying MDP exists, then the entropy regularized MDP exists \cite{lefebvre2022}. A sufficient condition for (\ref{eq:Qbullet}) is thus to verify that the reciprocal MDP exists.}
\end{document}